\documentclass[useAMS,usenatbib]{mn2e}
\usepackage[pdftex]{graphicx}
\usepackage[pdftex,usenames,dvipsnames]{xcolor}
\usepackage{natbib}
\usepackage{aas_macros}
\usepackage{url}
\usepackage[none]{hyphenat}
\usepackage{times}
\usepackage[linktocpage=true,backref=true,pagebackref=false,bookmarks=true,bookmarksnumbered=False,colorlinks=true,linkcolor=NavyBlue,citecolor=NavyBlue,urlcolor=NavyBlue]{hyperref}
\usepackage[all]{hypcap}
\pdfpageattr {/Group << /S /Transparency /I true /CS /DeviceRGB>>}  

\title[Radio observations of extreme ULXs]{Radio observations of extreme ULXs: revealing the most powerful ULX radio nebula ever or the jet of an intermediate-mass black hole?}

\author[M. Mezcua et al.]
  {M.~Mezcua,$^{1,2}$\thanks{Email: mmezcua@iac.es.}
  T.P.~Roberts,$^{3}$
  A.D.~Sutton,$^{3}$
  A.P.~Lobanov$^4$\thanks{Visiting Scientist, University of Hamburg / Deutsches Elektronen Synchrotron Forschungszentrum.}\\
  $^1$Instituto de Astrof\'isica de Canarias (IAC),
  		E-38200 La Laguna, Tenerife, Spain \\
  $^2$Universidad de La Laguna,
              Dept. Astrof\'isica, E-38206 La Laguna, Tenerife, Spain\\ 
  $^3$Department of Physics, University of Durham, South Road, Durham DH1 3LE, UK \\
  $^4$Max Planck Institute for Radio Astronomy,
              Auf dem H\"ugel 69, D-53121 Bonn, Germany }

\date{Accepted 2013 September 20}

\pagerange{\pageref{firstpage}--\pageref{lastpage}} \pubyear{2013}

\def\LaTeX{L\kern-.36em\raise.3ex\hbox{a}\kern-.15em
    T\kern-.1667em\lower.7ex\hbox{E}\kern-.125emX}

\begin{document}

\label{firstpage}

\maketitle

\begin{abstract}
The most extreme ultraluminous X-ray sources (ULXs), with $L_\mathrm{X} > 5 \times 10^{40}$ erg s$^{-1}$, are amongst the best candidates for hosting intermediate-mass black holes (IMBHs) in the haloes of galaxies. Jet radio emission is expected from a sub-Eddington accreting IMBH in the low/hard (radio bright) state. In a search for such IMBH jet radio emission, we have observed with the Very Large Array (VLA) at 5 GHz a sample of seven extreme ULXs whose X-ray properties indicate they are in
the hard state. Assuming they remain in this state, the non-detection of radio emission for six of the target sources allows us to constrain their black hole mass to the IMBH regime, thus ruling out a supermassive black hole nature. For the extreme ULX in the galaxy NGC\,2276, we detect extended radio emission formed by two lobes of total flux density 1.43 $\pm$ 0.22 mJy and size $\sim$650 pc. The X-ray counterpart is located between the two lobes, suggesting the presence of a black hole with jet radio emission. The radio luminosity allows us to constrain the black hole mass of this source to the IMBH regime; hence, the extreme ULX in NGC\,2276 could be the first detection of extended jet radio emission from an IMBH. The radio emission could also possibly come from a radio nebula powered by the ULX with a minimum total energy of $5.9 \times 10^{52}$ erg, thus constituting the most powerful and largest ULX radio nebula ever observed. 
 \end{abstract}

\begin{keywords}
accretion, accretion discs -- black hole physics -- ISM: jets and outflows -- Radio continuum: general -- X-rays: binaries.
\end{keywords}

\section{Introduction}
The search for intermediate-mass black holes (IMBHs), with masses between 100 and $10^{5} M\odot$ (e.g., \citealt{1999ApJ...519...89C}; \citealt{2004cbhg.symp...37V}), has gained increasing interest in the last decades as they constitute the missing link in studies of black hole (BH) growth and galaxy evolution. 
The presence of a supermassive black hole (SMBH) in the centre of (nearly) all galaxies (\citealt{1995ARA&A..33..581K}) is widely accepted. However, how these SMBHs form is not yet clearly understood. The finding of strong empirical correlations between the SMBH mass and their host galaxy properties (such as the bulge mass and stellar velocity dispersion; e.g., \citealt{2013ApJ...764..184M} and references therein) implies that the SMBH growth must be connected to the galaxy formation and/or evolution. These BH mass ($M_\mathrm{BH}$) scaling relationships predict the presence of IMBHs in the nucleus of low-mass galaxies, thus the presence of IMBHs is also expected in the haloes of large galaxies after tidal stripping of merging satellite galaxies (e.g., \citealt{2008ApJ...687L..57V}; \citealt{2010ApJ...721L.148B}).

IMBHs have also been invoked as the initial seed of SMBHs (e.g., \citealt{2010A&ARv..18..279V}; \citealt{2012ApJ...748L...7V}), which subsequently grow either through hierarchical BH merging or through gas accretion. 
IMBHs with masses $<10^{6} M\odot$ could form by the direct collapse of pre-galactic gas discs, while lower-mass IMBHs of $\sim200 M\odot$ could form from the collapse of a young and massive star (see \citealt{2012Sci...337..544V} and references therein) and IMBHs of masses $< 10^{3} M\odot$ could result from the collapse of an extremely massive star formed by the merging of massive young stars at the centre of a stellar cluster (e.g., \citealt{2004Natur.428..724P}). 
Such processes should also contribute to the existence of a population of IMBHs in the haloes of galaxies. Nevertheless, 
observational evidence of IMBHs is scarce.
Many studies have focused in seeking for IMBHs in low-luminosity active galactic nuclei (LLAGN) and in the centre of low-mass galaxies, where several IMBH candidates have been found (e.g., \citealt{2004ApJ...610..722G}; \citealt{2007ApJ...656...84G}; \citealt{2008AJ....136.1179B}; \citealt{2011Natur.470...66R}; \citealt{2012ApJ...753..103N}). 
In the haloes of galaxies, the most probable IMBHs candidates are the most extreme ultraluminous X-ray sources
(ULXs). ULXs are a type of extragalactic, non-nuclear X-ray sources whose isotropic luminosities $L_\mathrm{X}>10^{39}$ erg s$^{-1}$ in the 0.3-10 keV band are lower than the typical X-ray luminosities of AGN and brighter than those of Galactic X-ray binaries (XRBs; $M_\mathrm{BH} < 20 M\odot$) accreting at sub-Eddington rates. 
ULXs are typically associated with star-forming regions in nearby galaxies, and are sometimes embedded in optical and radio nebulae produced by the ULX radiative/kinetic energy output (e.g., \citealt{2004ApJS..154..519S, 2009ApJ...703..159S}; \citealt{2006IAUS..230..293P}; \citealt{2008AIPC.1010..303P}; \citealt{2011AN....332..371R}; \citealt{2012ApJ...749...17C}). Their high X-ray luminosities can be explained by XRBs of $M_\mathrm{BH}<100 M\odot$ with either super-Eddington accretion, anisotropic emission, and/or relativistic beaming (e.g., \citealt{2001ApJ...552L.109K}; \citealt{2002ApJ...568L..97B}; \citealt{2002A&A...382L..13K}; \citealt{2009MNRAS.397.1836G}; see reviews by \citealt{2007Ap&SS.311..203R}; \citealt{2011NewAR..55..166F}). Alternatively, ULXs could be also explained by sub-Eddington accreting IMBHs radiating isotropically, constituting thus a mass scaled-up version of XRBs. The later scenario was supported by X-ray spectral analysis of some ULXs, which revealed disc temperatures much cooler than those of XRBs but consistent with those expected from IMBHs accreting at $0.1L_\mathrm{Edd}$ rate (e.g., \citealt{2003ApJ...585L..37M, 2004ApJ...607..931M}; \citealt{2004MNRAS.355..359F}; \citealt{2007ApJ...660..580S}).
However, the presence of a cut-off at energies $>$ 3 keV and a soft excess in some of these ULX spectra, together with suppressed variability (e.g., \citealt{2006MNRAS.368..397S}; \citealt{2009MNRAS.397.1836G}; \citealt{2009MNRAS.397.1061H}), cannot be reconciled with sub-Eddington BHs but support the super-Eddington accretion scenario in which ULXs are XRBs in an ultraluminous state (\citealt{2009MNRAS.397.1836G}). Super-Eddington accretion, along with massive stellar BHs (up to $\sim100 M\odot$; \citealt{2009MNRAS.400..677Z}; \citealt{2010ApJ...714.1217B}) could explain ULXs up
to $\sim 10^{41}$ erg s$^{-1}$, but struggles to explain ULXs more luminous than this. Indeed, given that
the largest stellar BH mass is claimed at $\sim30 M_\odot$ (e.g., \citealt{2007ApJ...669L..21P}; \citealt{2008ApJ...678L..17S}) ULXs with X-ray luminosities in excess of $5 \times 10^{40}$ erg s$^{-1}$ cannot be
explained by analogues of known objects; hence, they constitute a compelling subset of targets
in the search for IMBHs.

Although observational evidence of these IMBHs is still scarce, a few strong candidates have been found based on X-ray timing and spectral analysis (e.g., M82 X-1, \citealt{2001ApJ...547L..25M}; Cartwheel N10, \citealt{2010MNRAS.406.1116P}; ESO 243-49 HLX-1, \citealt{2009Natur.460...73F}, \citealt{2012Sci...337..554W}). If these IMBH candidates behave as XRBs accreting at sub-Eddington rates, they are expected to follow spectral state transitions from a low/hard state in which steady, compact jet emission is observed in the radio regime to a high/soft state dominated by thermal disc emission.
Such X-ray transitions as well as variable radio emission from a transient jet were detected in ESO 243-49 HLX-1 (\citealt{2009ApJ...705L.109G}; \citealt{2011ApJ...743....6S}; \citealt{2012Sci...337..554W}); while steady jet emission from an IMBH has been suggested in a few sources (e.g., \citealt{2011AN....332..379M}; \citealt{2013arXiv1309.4463M}a)

X-ray analyses of a sample of extreme ULXs located in nearby spiral galaxies and closely
associated with star-forming regions revealed that the sources possess a hard power-law
continuum spectrum and display short-term variability (\citealt{2012MNRAS.423.1154S}). These properties
are consistent with sub-Eddington accreting BHs in the hard state, which given the observed
luminosities ($L_\mathrm{X} > 5 \times 10^{40}$ erg s$^{-1}$) suggest the presence of IMBHs with masses 
$10^{3} - 10^{5} M{\odot}$. If these sources are IMBHs in the hard state, they should be emitting steady radio jets.
With the aim of detecting radio emission (and possible IMBH steady jet emission) from these sources, we performed Very Large Array (VLA) radio observations at 5 GHz of seven of these extreme ULXs.
A description of this sample and the observations performed are presented in Section~\ref{observations}. The results obtained are described in Section~\ref{results} and discussed in Section~\ref{discussion}, while final conclusions are provided in Section~\ref{conclusions}.

\section{Sample and observations}
\label{observations}

\subsection{Sample}
The targets for the radio observations were drawn from the small sample of extreme ULXs studied by \cite{2012MNRAS.423.1154S}. In brief, these were selected from the ULX candidate catalogue of \cite{2011MNRAS.416.1844W} as the brightest ULX detections ($L_{\rm X} > 5 \times 10^{40} \rm ~erg~s^{-1}$ in the 0.3-10\,keV band) within 100 Mpc. Of the 10 candidates they initially selected, two were identified as background QSOs against an expectation of three background contaminants for the sample. The remaining eight objects were studied in detail, and revealed X-ray characteristics consistent with hard state IMBHs. For the purpose of this work, one object (Src. 7 of \citealt{2012MNRAS.423.1154S}, the brightest ULX candidate in the sample, hosted by the galaxy IC\,4320) was unfortunately too far south at declination of $-27^{\circ}$ to be observable by the VLA. The remaining seven objects therefore constituted our target sample.

\begin{table*}
\caption{The extreme ULX targets and observation details}
\label{table1}
\begin{center}
\begin{tabular}{cccccccc}
\hline
\hline
             &              &   		&\textit{Chandra}& \textit{Chandra}&               &                &                                   \\
ULX ID  & Host galaxy & Distance	& RA (J2000)  & 	Dec. (J2000)  & Obs. Date & Phase cal. & $S_\mathrm{5 GHz}$    \\
            &                    &     (Mpc)              &                       &                 &                 &             & (mJy beam$^{-1}$)      \\ 
\hline
Src. 1  	    &   NGC\,470	& 32.7	&  01:19:42.8  &  +03:24:22   &   May, 28  & J0125-0005 & $<0.012$ \\
Src. 2	    &   NGC\,1042     & 18.9	&  02:40:25.6   &  -08:24:30   &   May, 28  & J0241-0815 & $<0.017$ \\
Src. 3c	    &   NGC\,2276     & 33.3	&  07:26:48.3   &  +85:45:49  &   May, 6 \& 21 & J0508+8432 & 0.35 \\
Src. 5	    &   NGC\,4254     & 33.2	&  12:18:56.1   &  +14:24:19  &   May, 30  & J1224+2122 & $<0.017$ \\
Src. 8	    &   NGC\,5907     & 14.9	&  15:15:58.6   &  +56:18:10  &   May, 30  & J1438+6211 & $<0.020$ \\
Src. 9	    &   MCG\,11-20-19 & 96.2	&16:36:14.1   &  +66:14:10  &   May, 12  & J1642+6856 & $<0.062$ \\
Src. 10	    &   NGC\,7479     & 32.8	&  23:04:57.7   &  +12:20:29  &   May, 12  & J2250+1419 & $<0.061$ \\
\hline
\end{tabular}
\end{center}
{Column designation:}~ (1) ULX ID following the notation of \cite{2012MNRAS.423.1154S}; (2) host galaxy name; (3) host galaxy distance (from \citealt{2012MNRAS.423.1154S}); (4) right ascension at
the J2000 epoch; (5) declination at the J2000 epoch; (6) observation date in 2012 May; (7) phase calibrator; (8) peak flux density at the \textit{Chandra} position of each ULX.
\end{table*}

\subsection{VLA radio observations}
The seven target sources were observed with the NRAO EVLA\footnote{now officially known as the Karl G. Jansky VLA. The National Radio Astronomy Observatory is a facility of the National Science Foundation operated under cooperative agreement by Associated Universities, Inc.} (\citealt{2011ApJ...739L...1P}) at 5 GHz (project code 12A-183; PI: Mezcua). Three observing runs of $\sim$3 h in the B configuration were carried out for the sources in NGC\,470, NGC\,1042, NGC\,4254, NGC\,5907, MCG\,11-20-19 and NGC\,7479 in 2012 May (see Table~\ref{table1}). For NGC\,2276, two observing runs of 1.5 h were performed in the CnB configuration due to its high declination (+85$^{\circ}$). 
Each target was phase-referenced to a nearby (a few degrees away), bright, compact source (see Table~\ref{table1}), following a target-phase calibrator cycle of 9 min (7 min on the target source, 2 min on the phase calibrator). As a result, 1 h was spent on each target source. The sources 3C48 and 3C286 were used as flux calibrators to set the amplitude scale and as bandpass calibrators. The configuration of the WIDAR correlator was set to 16 spectral windows, each containing 128 channels of 1 MHz width. 

The correlated data were flagged, calibrated, and imaged using the Common Astronomy Software Applications (CASA) software. The presence of radio-frequency interference was minimal at the band of the observations. The flagging consisted of removing, for certain scans and spectral windows, a few channels and antennas which contained corrupted data. 
Data were calibrated in amplitude using the flux calibrators, while the phase calibrators were used to derive delay and gain solutions that were interpolated and applied to the target sources. For NGC\,2276, the calibrated data from its two observing runs were concatenated using the task CONCATENATE.
Each target was then imaged with CLEAN using the Cotton-Schwab algorithm and natural weighting. The resulting images have typical rms noises ranging 0.02--0.06 mJy beam$^{-1}$.

\section{Results and data analysis}
\label{results}
For the extreme ULXs in NGC\,470, NGC\,1042, NGC\,4254, NGC\,5907, MCG\,11-20-19 and NGC\,7479, no radio emission is detected above a 5$\sigma$ level in a 10 arcsec region around the \textit{Chandra} position of each ULX. Upper limits on the flux densities at 5 GHz of these objects, obtained from the local rms at their \textit{Chandra} positions, are provided in Table~\ref{table1} (Column 8).

\begin{figure}
  \includegraphics[width=\columnwidth]{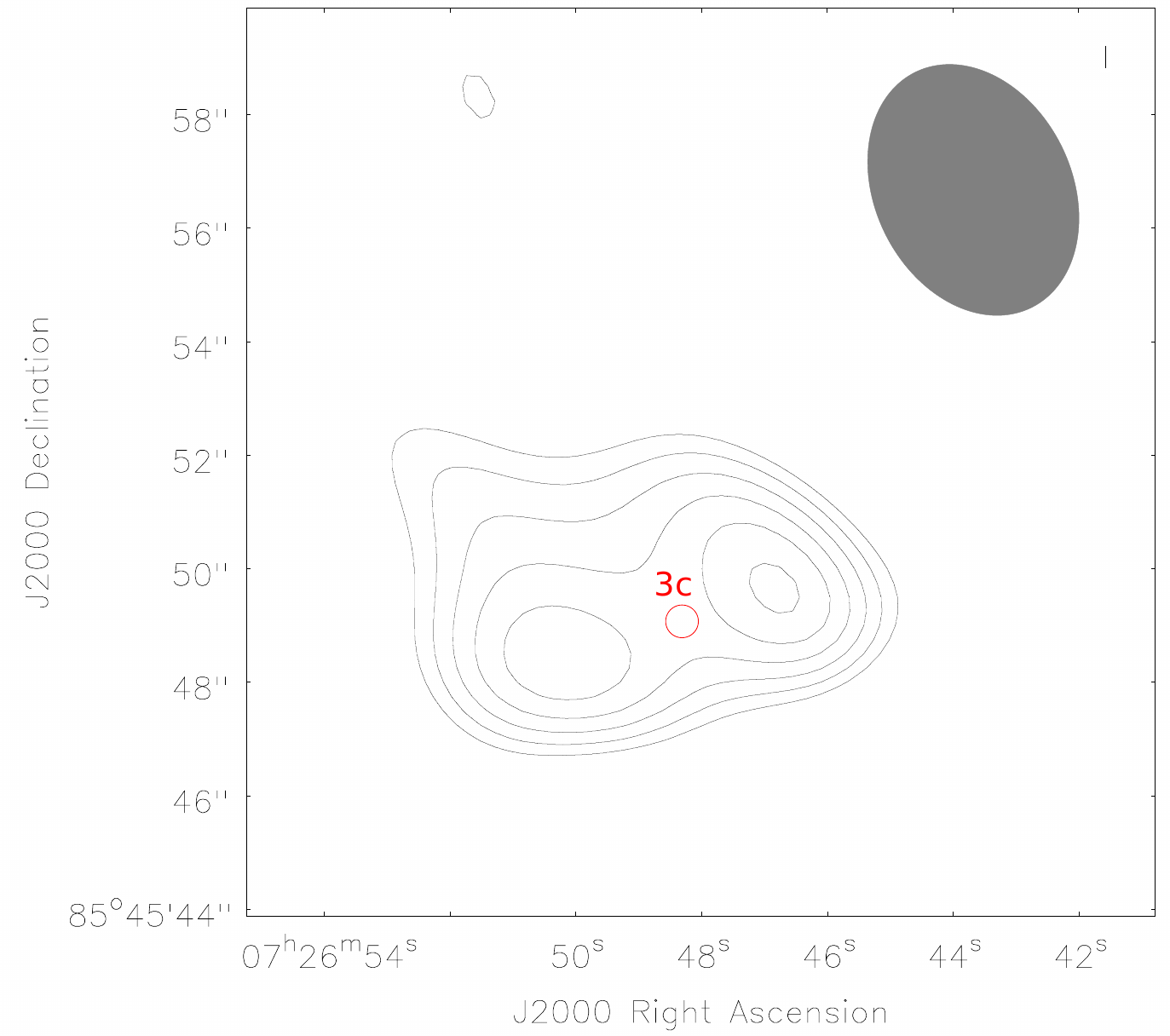}
\caption{VLA CnB-array image at 5 GHz of the extreme ULX in NGC\,2276. The contours are (5, 6, 7, 8, 9, 10) times the off-source rms noise of 0.04 mJy beam$^{-1}$. The beam size is 4.6 $\times$ 3.5 arcsec$^{2}$ at a position angle of $26^{\circ}$. The \textit{Chandra} position of the source 3c is marked with a red circle of diameter 0.6 arcsec.}
\label{fig1}
\end{figure}

For the extreme ULX 3c in NGC\,2276, extended radio emission is detected around its \textit{Chandra} position (see Fig.~\ref{fig1}). The radio structure has an integrated flux density of 1.43 $\pm$ 0.22 mJy, corresponding to an integrated luminosity of 9.51 $\times$ 10$^{36}$ erg s$^{-1}$ at the distance of the galaxy, and is formed by two lobes oriented NW-SE. Both peaks of emission are detected at a $\sim10\sigma$ level, and the distance between them is of $\sim4$ arcsec ($\sim650$ pc at the distance of NGC 2276). The size measured from the lowest contour ($5\sigma$) on the image plane is 8.5 $\times$ 5.3 arcsec$^{2}$, which corresponds to 1372.3 $\times$ 855.7 pc at the distance of the galaxy.
After exporting the cleaned image to AIPS\footnote{Astronomical Image Processing Software of NRAO.}, we fitted the western (corresponding to the main peak of emission) and the eastern lobe with a double two-dimensional elliptical Gaussian using the AIPS task JMFIT. This yields a peak intensity\footnote{the error on the flux comes from a combination of the rms noise with the uncertainty provided by the fit of the Gaussian component.} of 0.35 $\pm$ 0.06 mJy beam$^{-1}$ centred at RA(J2000) = 07$^h$26$^m$46$^s$.77 $\pm$ 0$^s$.28, 
Dec.(J2000) = 85$^{\circ}$45\arcmin49\arcsec.75 $\pm$ 0\arcsec.47. The peak intensity obtained for the eastern lobe is of 0.34 $\pm$ 0.06 mJy beam$^{-1}$, centred at RA(J2000) = 07$^h$26$^m$50$^s$.16 $\pm$ 0$^s$.34, 
Dec.(J2000) = 85$^{\circ}$45\arcmin48\arcsec.50 $\pm$ 0\arcsec.78.
The \textit{Chandra} source 3c is located at the centre of the extended radio structure, in the middle of the two lobes, and is offset by 1.8 arcsec to the main peak of radio emission.
The brightness of the two lobes is essentially the same, which indicates that
either the emission is not Doppler boosted or the velocity vector of the emitting material lies very close to the picture plane. Distinguishing between
these two possibilities requires imaging the structure of the radio emission at a higher resolution with Very Long Baseline Interferometry (VLBI).

Robust weighting (\citealt{1995AAS...18711202B}) was also used to image NGC\,2276 in order to check if any further components could be resolved. This weighting scheme yielded the detection of the same two components, though with a much lower signal-to-noise ratio (S/N$\sim$5).

\section{Discussion}
\label{discussion}
\subsection{The nature of the extreme ULXs without radio counterparts}
\label{non-detections}
The X-ray spectra of all those extreme ULXs with non-detections of radio emission are all well fitted by absorbed power-law continua with hard photon indexes ($\Gamma = 1.4 - 2.2$) in at least one epoch. Most also display at least one detection of significant variability (fractional variability $\sim$10\%; \citealt{2012MNRAS.423.1154S}). 
These characteristics are consistent with the low/hard state, usually described by a power-law spectrum of $\Gamma \sim 1.7$ and fractional variability $\sim$10\% (\citealt{2006ARA&A..44...49R}), typically seen in Galactic XRBs accreting at $\leq0.1$ Eddington rate.  
Assuming that the extreme ULXs are accreting at 10\% the Eddington limit, lower limits on the BH mass in the range $(1.9-6.9) \times 10^{3} M{\odot}$ were calculated for the seven sources in the radio sample (\citealt{2012MNRAS.423.1154S}), which imply the presence of IMBHs in these ULXs.

For those extreme ULXs in the low/hard state whose spectrum is well described by an absorbed power-law model, an estimate of their $M_\mathrm{BH}$ can be derived from the X-ray-radio Fundamental Plane of accreting BHs (e.g., \citealt{2003MNRAS.345.1057M}; \citealt{2004A&A...414..895F}; \citealt{2006A&A...456..439K}). Upper limits on the $M_\mathrm{BH}$ were already obtained by \cite{2012MNRAS.423.1154S} using the completeness limits of the FIRST\footnote{Faint Images of the Radio Sky at Twenty-cm.} (\citealt{1995ApJ...450..559B}) and NVSS\footnote{NRAO VLA Sky Survey} (\citealt{1998AJ....115.1693C}) catalogues as an upper limit to the radio fluxes (2.5 mJy and 1 mJy, respectively). Flat radio spectral indexes were assumed to scale the flux densities from 1.4 to 5 GHz. 
From our VLA non-detections at 5 GHz we can improve the mass constraints without any assumption on the spectral index. The lack of simultaneity of the radio and X-ray observations will constitute the major source of uncertainty in the mass estimates.  
However, most of the extreme ULXs vary by no more than a 
factor of a few ($\sim$2--5) on a baseline of 1--10 years (cf. fig.~8 of \citealt{2012MNRAS.423.1154S}), so
we do not expect them to fade dramatically before the radio observations.
Converting the highest X-ray fluxes in the 0.3--10 keV band to the 2--10 keV band using WebPIMMS\footnote{http://heasarc.nasa.gov/Tools/w3pimms.html} and calculating the X-ray and radio luminosities at the distances of each ULX host galaxy (see Table 1), we estimate mass upper limits in the range $3.5 \times 10^{3} - 1.4 \times 10^{6} M{\odot}$ (see Table~\ref{table2}) using the X-ray/radio Fundamental Plane from \cite{2003MNRAS.345.1057M}\footnote{These mass upper limits can vary depending on the correlation used (using the fundamental place of \citealt{2006A&A...456..439K} 
the mass upper limits increase by a factor of $\sim$2--8, still constraining the masses of all ULXs except for two to the IMBH regime.}. These mass limits represent an improvement of $\sim$1--2.5 orders of magnitude with respect to the mass upper limits of \cite{2012MNRAS.423.1154S} except for source 3c, where the detection of the extended nebula/jet potentially masks the fainter radio core. The new data not only place better mass limits but also exclude SMBHs in all but one case (Src. 9) and constrain the masses to the IMBH regime.
It should be noted that some of the limits are getting very close together - in the case of Src. 8 they almost touch (at 
$3.2 - 3.5 \times 10^{3} M{\odot}$, but, as discussed below, we think this source is a super-Eddington stellar BH).

One alternative possibility that might explain the lack of radio emission in most of the ULXs is that the sources might have transitioned from the low/hard state to a soft/high state. These state transitions are usually seen in sub-Eddington accreting XRBs, whose X-ray spectrum is better described by an absorbed multi-colour disc blackbody (MCD) model at their soft X-ray peak luminosity and by an absorbed power-law model at lower luminosities. These have also been suggested in the IMBH candidates ESO 243-49 HLX-1 (e.g., \citealt{2011ApJ...743....6S}), M82-X1 (\citealt{2010ApJ...712L.169F}) and Src. 1 in NGC\,470 (\citealt{2012MNRAS.423.1154S}), though the latter two are questionable identifications since the disc spectra for Src. 1 and M82-X1 are far too hot ($\sim$1.5 -- 2 keV) for an IMBH.

However, we do not expect all the sources to switch to the thermal state simultaneously. A further alternative scenario for the non-detection of radio emission is that the sources might be super-Eddington. This can be the case of Src. 8 in NGC\,5907, highlighted by \cite{2012MNRAS.423.1154S} as being the only object in their sample to display an X-ray spectral break at energies of $\sim$5 keV indicative of the super-Eddington ultraluminous state. Interestingly, further study of this ULX based on new {\it XMM-Newton, Chandra\/} and {\it Swift\/} observations (Sutton et al. in preparation) suggests that the hard spectrum and variability of this object (that are quantitatively similar to the lower-quality observations of the other objects) may be due to a close to face-on viewing angle and its high accretion rate. This could lead to the super-Eddington radiatively driven wind becoming so extensive that its temperature drops to the point where a combination of the low temperature and a high absorption column effectively hide it. Given the high accretion rate, though, the wind is close to spherical so its clumpy edge can impinge on the line of sight to the spectrally hard central regions, imprinting extrinsic variability as described by \cite{2011MNRAS.411..644M} on to the geometrically beamed hard emission from close to the BH. We note that this combination of mechanisms could potentially explain the high luminosities, hard spectra, and variability observed in the X-rays for the whole sample of extreme ULXs, thus providing an alternative explanation for the lack of radio emission, although there may still be some necessity for large stellar BHs in this scenario (e.g., \citealt{2009MNRAS.400..677Z}) due to the extreme X-ray luminosities observed.

\subsection{The extreme ULX in NGC\,2276}
\subsubsection{The radio jet of an IMBH?}
The initial \textit{XMM-Newton} detection of the extreme ULX in NGC\,2276 was resolved into a triplet of sources by \textit{Chandra}, with objects b \& c closest to the \textit{XMM} centroid (\citealt{2012MNRAS.423.1154S}, Fig. 1). The 5 GHz VLA radio observations of the source 3c in NGC\,2276 reveal a
counterpart of total flux density 1.43 mJy and luminosity $L_\mathrm{5 GHz}=9.51 \times 10^{36}$ erg s$^{-1}$.
The ULX radio counterpart is extended, looking like two sources of
size $\sim$650 pc, and the X-ray
source 3c is in the middle of them (Fig.~\ref{fig1}). The structure is thus reminiscent of two radio
lobes powered by a central BH. 

In the \textit{Chandra} X-ray observations (\citealt{2012MNRAS.423.1154S}) the spectrum of Src. 3c was fitted by an absorbed power-law spectrum with $\Gamma=1.7$, yielding a luminosity $L_\mathrm{0.3-10 keV}\approx 5.2 \times 10^{39}$ erg s$^{-1}$. The blended triplet had a much higher peak luminosity of $L_\mathrm{0.2-12 keV}\approx 6.4 \times 10^{40}$ erg s$^{-1}$ and its spectrum was also well fitted by a power-law model with $\Gamma=1.4$ (\citealt{2012MNRAS.423.1154S}). 
The X-ray spectral and timing properties of the source provided by these observations
suggest that it could host an IMBH in the low/hard state, with a BH mass $\geq 4700 M{\odot}$ estimated using the \textit{XMM} peak luminosity and assuming 10\% Eddington accretion (\citealt{2012MNRAS.423.1154S}). 
In this case, the two lobes of radio emission could be the first ever evidence of extended jet emission from an IMBH. To date very few cases of transient and steady jet radio emission from an IMBH have been reported (\citealt{2011AN....332..379M}; \citealt{2012Sci...337..554W}; \citealt{2013arXiv1309.4463M}a), but in none of them is the structure of the putative radio jet resolved.  The most powerful extragalactic non-nuclear radio jet ever discovered belongs to the microquasar S26 in NGC\,7793 (which is 300 pc across and has a flux density $\sim$2 mJy; \citealt{2010Natur.466..209P}; \citealt{2010MNRAS.409..541S}), while the most powerful example in our own Galaxy corresponds to SS433 (e.g., \citealt{2006MNRAS.370..399B}).
With a size of $\sim$650 pc, the IMBH jet candidate in NGC\,2276 would thus be the largest non-nuclear jet ever detected. 

The detection of a jet core would allow us to test the BH scaling relations in the intermediate-mass regime.
The $M_\mathrm{BH}$ of this IMBH candidate could be constrained via the Fundamental Plane correlation if compact radio emission were detected within the error circle of the \textit{Chandra} position of the ULX. The non-detection of a compact radio component does not allow such a mass constraint. However, an upper limit on the $M_\mathrm{BH}$ can be estimated from the radio flux density of the main peak of emission and scaling the 0.2--12 keV flux to the 2-10 keV band (as performed in Section~\ref{non-detections}). This yields $M_\mathrm{BH}\leq 8.5 \times 10^{5} M{\odot}$.
Under the assumption of 10\% Eddington accretion, the ULX $M_\mathrm{BH}$ is thus constrained to $4700 \leq M_\mathrm{BH}\leq 8.5 \times 10^{5} M\odot$, ruling out a SMBH nature. The major caveat of this mass range is the lack of simultaneity between the radio and X-ray observations. To solve this issue and with the aim of resolving the putative jet emission and detecting a compact core component, we have been granted European VLBI Network (EVN\footnote{www.evlbi.org}) observations simultaneous with \textit{Chandra} X-ray observations of this source.

\begin{table}
\caption{BH mass limits}
\label{table2}
\begin{center}
\begin{tabular}{ccccccc}
\hline
\hline
ULX ID  & $M_\mathrm{BH} (M{\odot})\ ^{a}$  &       $M_\mathrm{BH} (M{\odot})\ ^{b}$  \\
            &                                                             &                               				\\ 
\hline
Src. 1  	    &   $\geq1.9 \times 10^{3} $                &			$\leq8.5 \times 10^{3}$	\\
Src. 2	    &   $\geq3.5 \times 10^{3}$                 &			$\leq5.4 \times 10^{3}$	\\
Src. 3c	    &   $\geq4.7 \times 10^{3}$                 &		       $\leq8.5 \times 10^{5}$ 	\\
Src. 5	    &   $\geq6.9 \times 10^{3}$                 &			$\leq1.4 \times 10^{4}$	\\
Src. 8	    &   $\geq3.2 \times 10^{3}$                 &			$\leq3.5 \times 10^{3}$	\\
Src. 9	    &   $\geq5.2 \times 10^{3}$                 &			$\leq1.4 \times 10^{6}$	\\
Src. 10	    &  $\geq4.7 \times 10^{3}$                  &			$\leq9.3 \times 10^{4}$	\\
\hline
\end{tabular}
\end{center}
$^{a}$ BH mass assuming a hard state with power-law spectrum, maximum luminosity, and accretion $\leq$10\% Eddington rate (from \citealt{2012MNRAS.423.1154S}). $^{b}$ BH mass estimated from the Fundamental Plane of \cite{2003MNRAS.345.1057M}.
\end{table}

\subsubsection{A radio nebula around an IMBH?}
In \textit{F606W} and \textit{F814W} \textit{Hubble Space Telescope} (\textit{HST}) images, the extreme ULX 3c seems to lie close to a star-forming region in the spiral arm of NGC\,2276 (\citealt{2012MNRAS.423.1154S}).
A candidate optical counterpart of absolute magnitude $\sim-9$ was reported by these authors, which is consistent with the presence of a young massive stellar cluster and would confirm the association of the source with a star-forming complex. This would provide a natural explanation on how the IMBH was formed, as one of the proposed IMBH formation mechanisms is the collapse of a massive star cluster (e.g., \citealt{2004Natur.428..724P}). On the other hand, NGC\,2276 is located in a small group of galaxies around the central elliptical galaxy NGC\,2300 and is suggested to be interacting, either tidally with NGC\,2300, or with the hot intergroup gas (e.g., \citealt{1993ApJ...413L..81G}; \citealt{1991A&A...244...52E}; \citealt{1996ApJ...460..601D}). This would explain the truncated shape of the galaxy and the enhancement of star formation observed on one side of it (e.g., \citealt{1997AJ....114..613D}; cf. our Fig.~\ref{fig4}). The IMBH would thus most probably have been formed in situ in the star-forming region, although the possibility that the source is the nucleus of a captured satellite galaxy cannot be ruled out.

\begin{figure}
  \includegraphics[width=\columnwidth]{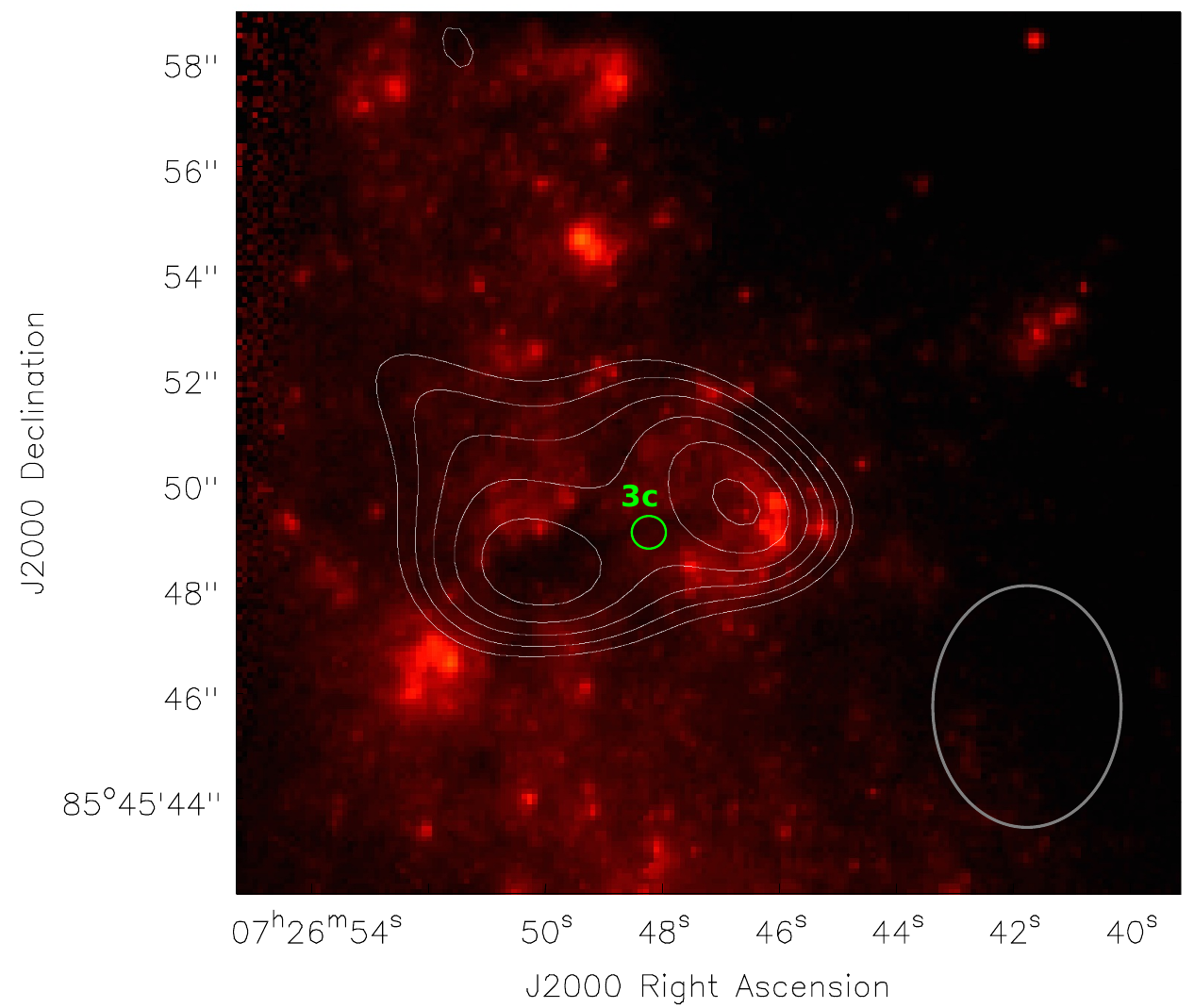}
\caption{\textit{F606W} \textit{HST} image of the extreme ULX 3c in NGC\,2276. The VLA CnB-array radio contours at 5 GHz are plotted as (5, 6, 7, 8, 9, 10) times the off-source rms noise of 0.04 mJy beam$^{-1}$. 
The size of the VLA beam (4.6 $\times$ 3.5 arcsec$^{2}$) is indicated with a grey ellipse. 
The green circle denotes the \textit{Chandra}
position of 3c with a diameter of 0.6 arcsec. North is up, East to the left.}
\label{fig3}
\end{figure}

\begin{figure*}
  \includegraphics[width=\textwidth]{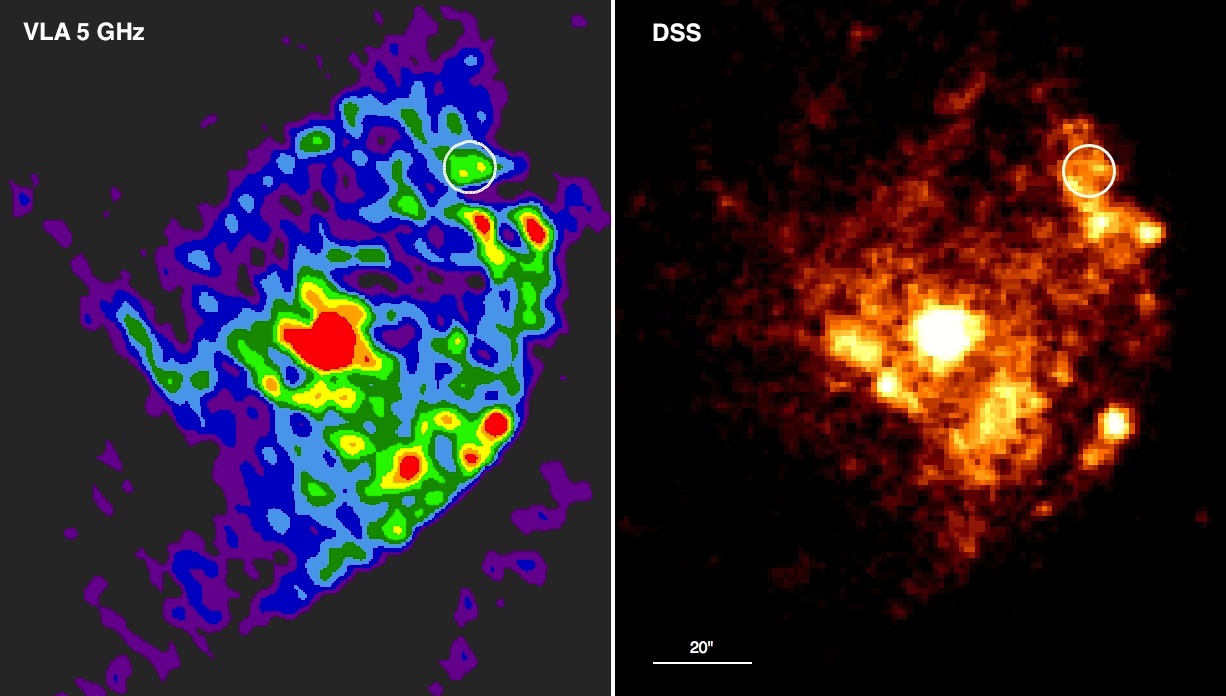}
\caption{Left: VLA CnB-array image at 5 GHz of NGC\,2276. Same rms noise and beam size as in Fig.~\ref{fig1}. Right: Digitized Sky Survey (\citealt{1990AJ.....99.2019L}) image of NGC\,2276. The location of the ULX is marked with a white circle of radius 5 arcsec.
The field size is 2 arcmin $\times$ 2 arcmin.}
\label{fig4}
\end{figure*}

In a star-forming environment, the extended radio structure detected with the VLA could correspond to either jet radio emission from the IMBH or a radio nebula powered by the IMBH. The superposition of the radio image on the \textit{HST} \textit{F606W} image (Fig.~\ref{fig3}) reveals clumps of star formation or star clusters lying very close to the edges of the radio contours, but this does not allow ruling out any of the two scenarios. 
The minimum total energy of the radio jet (or to power the radio nebula) can be calculated assuming synchrotron radiation, equipartition condition between the energy of the radiating particles and the magnetic field, and excluding relativistic protons (e.g., \citealt{1994hea..book.....L}). 
Taking a prolate spheroid as the volume of the emitting region, $V=\frac{\pi}{6}d_\mathrm{maj}d_\mathrm{min}^{2}$ with $d_\mathrm{maj}=8.5\arcsec$ and $d_\mathrm{min}=5.3\arcsec$, yields $V=1.54 \times 10^{58}$ m$^{3}$. Using this volume, a filling factor of unity, and $L_{\nu}=1.90 \times 10^{20}$ W Hz$^{-1}$, we find that the minimum total energy is $E_\mathrm{min}=5.94 \times 10^{52}$ erg with a magnetic field strength of $6.5 \mu$G, and that the synchrotron lifetime of the electrons is 23 Myr. 
While the electron lifetime is similar to those found in other ULXs surrounded by radio nebulae (e.g., 19 Myr for IC\,342\,X1, \citealt{2012ApJ...749...17C}; 20 Myr for NGC\,5408\,X1, \citealt{2007ApJ...666...79L}; 25 Myr for Ho\,II\,X-1, \citealt{2005ApJ...623L.109M}), the minimum energy required to power the putative radio nebula of NGC\,2276 is a factor of 65 higher than that of IC\,342\,X1 ($E_\mathrm{min}= 9.18 \times 10^{50}$ erg; \citealt{2012ApJ...749...17C}), $\sim1.7 \times 10^{3}$ times higher than for NGC\,5408\,X1 ($E_\mathrm{min}= 3.6 \times 10^{49}$ erg; \citealt{2007ApJ...666...79L}), and $\sim2.3 \times 10^{3}$ times higher than for Ho\,II\,X1 ($E_\mathrm{min}= 2.6 \times 10^{49}$ erg; \citealt{2005ApJ...623L.109M}). The high $E_\mathrm{min}$ derived for Src. 3 implies that the ULX is required to have been energising the nebula for 200,000 years at a rate of 10$^{40}$ erg s$^{-1}$ (presumably kinetic energy, in addition to its radiative emission) to create the nebula.

The size of NGC\,2276\,3c is also much larger than these nebulae (approximately five times larger than IC\,342\,X1, and $\sim18$ times larger than NGC\,5408\,X1 and Ho\,II\,X1 when considering a size of 650 pc for NGC\,2276\,3c). Therefore, if the radio structure found in 3c belonged to a radio nebula, it would be the largest and most powerful ULX radio nebula ever observed. 
The EVN radio observations will allow us to confirm/reject this scenario by resolving the VLA radio emission of this extreme ULX. 

Finally, it should be also mentioned the possibility that Src. 3 is a background radio-loud AGN, a foreground source, or a chance alignment of the X-ray emission with radio emission in the galaxy (e.g., from SNe). The former can be tested by deriving the chance alignment between the VLA and the \textit{Chandra} counterparts. Using the number of detected sources with the same or greater flux as the VLA radio counterpart (12 sources within a field of view of 4 arcmin) and the \textit{Chandra} PSF full width at half maximum of 0.5 arcsec, we obtain a probability of change alignment of 0.03\%, which is quite low. In addition, the alignment of the X-ray source to the radio lobes and optical emission (Fig.~\ref{fig3}) strongly suggests that they are all linked, hence making the chance alignment scenario very unlikely. This radio-optical-X-ray alignment together with the observation that both the optical and radio emission certainly seem to belong to the spiral arms of the host galaxy (see Fig.~\ref{fig4}), potentially indicate that Src. 3 must belong to NGC\,2276 and make the background source scenario also very unlikely. 
On the other hand, the high column density and lack of a bright counterpart in the \textit{HST} images strongly exclude a Galactic foreground object as the source of the radio and X-ray emission. 


\section{Conclusions}
\label{conclusions}
IMBHs have been long predicted from $M_\mathrm{BH}$ scaling relationships and their existence might be key to explaining the growth of the most massive BHs in the centre of galaxies (e.g., \citealt{2008ApJ...687L..57V}; \citealt{2012ApJ...748L...7V}). The formation of IMBHs has been suggested to result from the collapse of massive stars in stellar clusters or the collapse of pre-galactic gas disc (e.g., \citealt{2012Sci...337..544V}), hence a population of IMBHs should be present in the haloes of galaxies. 
Extreme ULXs, with X-ray luminosities in excess of $5 \times 10^{40}$ erg s$^{-1}$, are amongst the best IMBH candidates in the haloes of galaxies, as it is difficult to explain their high luminosities using super-Eddington emission and beaming
scenarios for known stellar BH masses. X-ray analyses of a sample of extreme ULXs (\citealt{2012MNRAS.423.1154S}) revealed properties such as hard X-ray spectra and short-term X-ray variability, which make these objects potential candidates for IMBHs in the low/hard state. Jet radio emission from a sub-Eddington accreting BH is expected in this state, in which an estimate of the $M_\mathrm{BH}$ can be obtained from the radio/X-ray correlation or Fundamental Plane of accreting BHs. Obtaining an estimate of the $M_\mathrm{BH}$ of these sources is thus crucial to support the IMBH nature scenario and reveal the presence of a radio jet. With this aim, we observed a sample of extreme ULXs with the VLA at 5 GHz. Although no radio emission was detected for six out of the seven target sources, the upper limits on the radio flux densities (assuming they remain in the hard state) allow us to constrain the $M_\mathrm{BH}$ of almost all sources to the IMBH regime, ruling out their possible nature as SMBHs. Radio observations are thus crucial in order to reveal the nature of these sources. 
For the source Src. 3 in NGC\,2276, extended radio emission of flux density 1.43 mJy and size $\sim$650 pc is detected. The emission is consistent with the \textit{Chandra} X-ray position, which lies in between the two observed lobes of radio emission and thus suggests the presence of a central BH. Using the radio/X-ray correlation, 
the $M_\mathrm{BH}$ of the source is constrained to $4.7 \times 10^{3} \leq M_\mathrm{BH} \leq 8.5 \times 10^{5} M{\odot}$, which indicates that the radio emission might be coming from the jet of an IMBH. Alternatively, the ULX could be powering a radio nebula with $E_\mathrm{min}=5.9 \times 10^{52}$ erg. Follow-up VLBI radio observations will allow us to reveal the nature of the radio emission.

\section*{Acknowledgements}
The authors are grateful to the insightful suggestions of the referee, which have helped to improve the manuscript.
TPR acknowledges support from the UK STFC in the form of the standard grant ST/G001588/1 and as part of the consolidated grant ST/K000861/1, and ADS was funded by STFC via a PhD studentship (ST/F007299/1).
Based on observations made with the NASA/ESA Hubble Space Telescope, obtained from the data archive at the Space Telescope Science Institute. STScI is operated by the Association of Universities for Research in Astronomy, Inc. under NASA contract NAS 5-26555.

\bibliographystyle{mn2e} 
\bibliography{referencesALL}

\label{lastpage}

\end{document}